\documentstyle[12pt]{article}

\advance \textheight by \topskip
\title{Model of Large Mixing Angle MSW Solution}

\author{Morimitsu TANIMOTO
 \\
Science Education Laboratory, Ehime University,
 \\ Matsuyama, Japan, 790-8577}

\def\bar{\overline}

\def\d{\delta}
\def\a{\alpha}
\def\b{\beta}
\def\n{\nu}
\def\m{\mu}

\def\e{\epsilon}

\def\th{\theta}

\def\bar{\overline}

\def\eV{{\rm eV}}

\begin{document}
\maketitle

\begin{abstract}

 We have obtained the neutrino mass matrix with the large mixing angle
(LMA) MSW 
 solution, 
 $\sin^2 2\th_\odot=0.65\sim 0.97$ and $\Delta m_{\odot}^2= 10^{-5}\sim
10^{-4}\eV^2$,
   in the $S_{3L}\times S_{3R}$ flavor symmetry.
 The structure of our neutrino mass matrix is found to be 
 stable against radiative corrections. 
\end{abstract}
 

   The solar neutrino data as well as the atmospheric one 
	 give big impact on the study of the lepton mass matrices. 
   There is a typical texture of the lepton mass matrix with
	  the nearly maximal mixing of flavors, which
	 is derived from the symmetry of the lepton flavor democracy \cite{FX,Lisbon}, 
	 or  from the $S_{3L}\times S_{3R}$ symmetry 
	 of the left-handed Majorana neutrino mass matrix \cite{FTY,KK}. 
	 This texture have given a prediction for the neutrino mixing 
	  $\sin^2 2\th_{\rm atm}=8/9$. The mixing for the solar neutrino
	  depends on the symmetry breaking pattern of the flavor such as
	  $\sin^2 2\th_\odot= 1$ or $\ll 1$.
  However, the LMA-MSW solution, $\sin^2 2\th_\odot=0.65\sim 0.97$ and
	$\Delta m_{\odot}^2= 10^{-5}\sim 10^{-4}\eV^2$ \cite{BKS},  
	  has not been obtained in the previous works \cite{FX,Lisbon,FTY,KK}.
 We study how to get the LMA-MSW solution
   in the $S_{3L}\times S_{3R}$ symmetric mass matrices, 
   and discuss the stability of the neutrino mass matrix against radiative 
  corrections.


The texture of the charged lepton mass matrix
was presented based on the $S_{3L}\times S_{3R}$ symmetry
  as follows \cite{FX,FTY,Koide}:
\begin{equation}
M_\ell= {c_\ell \over 3}
             \left (\matrix{1 & 1 & 1 \cr
			                1 & 1 & 1 \cr
                            1 & 1 & 1 \cr  } \right)  + M_\ell^{(c)}\ ,
\label{MC}
\end{equation}
\noindent where the second matrix is the flavor symmetry breaking one.
The unitary matrix $V_\ell$, which diagonalizes the mass matrix $M_\ell$,
 is given as $V_\ell=F L$, where
\begin{equation}
F=\left( \matrix{1/\sqrt 2 & 1/\sqrt 6 & 1/\sqrt 3 \cr
                   -1/\sqrt 2 & 1/\sqrt 6 & 1/\sqrt 3 \cr
                           0 & -2/\sqrt 6 & 1/\sqrt 3 \cr } \right) 
\label{F} 
\end{equation}										 
\noindent diagonalizes the democratic matrix and $L$ depends on
the mass correction term $M_\ell^{(c)}$.

 The neutrino mass matrix is different from the democratic one 
 if they are Majorana particles.
The $S_{3L}$ symmetric mass term is given as follows:
\begin{equation}
   {c_\n}  \left( \matrix{1 & 0 & 0 \cr
                            0 & 1 & 0 \cr
                            0 & 0 & 1 \cr  } \right)
	+ {c_\n} r \left( \matrix{1 & 1 & 1 \cr
                            1 & 1 & 1 \cr
                            1 & 1 & 1 \cr  } \right) \ , 
  \end{equation}
\noindent
where $c_\n$ and $r$ are arbitrary parameters.
The eigenvalues of this matrix are easily obtained by using the orthogonal
matrix
 $F$ in eq.(\ref{F}) as 
$c_\n(1, \ 1, \ 1+3r)$.
 
 The simplest breaking terms of the $S_{3L}$ symmetry are added in (3,3)
 and (2,2) entries.  Therefore, the neutrino mass matrix is written as
 \begin{equation}
 M_\n= {c_\n}  \left( \matrix{1 + r & r & r \cr
                              r & 1+r+\e & r \cr
                              r & r & 1+r+\d \cr  } \right) \ ,
 \label{Mass}
 \end{equation}
 \noindent
 in terms of small breaking parameters $\e$ and $\d$.
 In order to explain both solar and atmospheric neutrinos in this mass matrix,
  $r\ll 1$ should be satisfied. 
  However, there is no reason why $r$ is very small in this framework.
  In order to answer this question,  we need a higher symmetry 
  of  flavors such as  the $O_{3L}\times O_{3R}$ model \cite{O3}.

Let us consider the case of 
   $\d\gg \e \simeq r$, in which  $S_{3L}$ symmetry is completely broken.
  Then neutrino mass eigenvalues are given as
\begin{equation}
  m_1 \simeq 1 + \frac{1}{2}\e+ r - \frac{1}{2}\sqrt{\e^2+4 r^2} \ , \quad 
  m_2 \simeq 1 + \frac{1}{2}\e+ r + \frac{1}{2}\sqrt{\e^2+4 r^2}\ , 
  \quad   m_3 \simeq 1 + r + \d ,
\end{equation}
\noindent in the $c_\n$ unit.
The orthogonal matrix $U_\n$ is given as
 \begin{equation}
U_\n \simeq \left (\matrix{t & \sqrt{1-t^2} & \frac{r}{\d} \cr
			         -\sqrt{1-t^2} & t & \frac{r}{\d-\e} \cr
  \frac{r}{\d}(\sqrt{1-t^2}-t) & -\frac{r}{\d-\e}(t+\sqrt{1-t^2}) & 1
\cr}\right)\ ,
  \quad t^2=\frac{1}{2}+\frac{1}{2}\frac{\e}{\sqrt{\e^2+4r^2}} \ . 
\label{Mix2}
\end{equation}

 Since the correction term $L$ is close to the unit matrix, 
the MNS matrix $U_{\a i}$ is approximately given as
  $F^T U_\n$ as follows:
 
  \begin{equation}
   F^T U_\n \simeq 
   \left (\matrix{\frac{1}{\sqrt{2}}(t+\sqrt{1-t^2})
 &\frac{1}{\sqrt{2}}(\sqrt{1-t^2}-t) & -\frac{1}{\sqrt{2}}\frac{\e
r}{\d(\d-\e)} \cr
	\frac{1}{\sqrt{6}}(t-\sqrt{1-t^2})(1+\frac{2r}{\d})
	 & \frac{1}{\sqrt{6}}(t+\sqrt{1-t^2})(1+\frac{2r}{\d-\e})
	 & -\frac{2}{\sqrt{6}}(1-\frac{r}{\d}) \cr
  \frac{1}{\sqrt{3}}(t-\sqrt{1-t^2})(1-\frac{r}{\d})
   &  \frac{1}{\sqrt{3}}(t+\sqrt{1-t^2})(1-\frac{r}{\d-\e}) 
   & \frac{1}{\sqrt{3}}(1+\frac{2r}{\d}) \cr}\right)\ .
  \end{equation}
  
  \noindent
  The mixing angle between the first and second flavor depends on $t$, which
  is determined by $r/\e$. 
   It becomes the maximal angle in the case of $t=1$ ($r/\e=0$) and
   the minimal  one in the case of $t=1/\sqrt{2}$ ($\e/r=0$).
   Since we get $\sin^2 2\th_\odot=\e^2/(\e^2+4 r^2)$, 
    the relevant value of  $r/\e$ leads easily to 
    $\sin^2 2\th_\odot=0.65\sim 0.97$, which corresponds to the LMA-MSW
solution.
  The numerical results have been shown in ref. \cite{tanimoto}.
  
  We should carefully discuss the stability of our results against radiative 
  corrections since the model predicts nearly degenerate neutrinos.
  When the texture of the mass matrix is given 
  at the $S_{3L}\times S_{3R}$ symmetry energy scale, radiative corrections
are not
  negligible at the electoroweak (EW) scale.
  
  Let us consider the basis, in which the mass matrix of the charged leptons
  is diagonal.  The neutrino mass matrix in eq.(\ref{Mass}) is transformed into
  $V_\ell^\dagger M_\n V_\ell$.
 The radiatively corrected mass matrix in the MSSM at the EW scale is given 
 as $R_G \bar M_\n R_G$, where $R_G$ is given by RGE's \cite{Haba} as
  \begin{equation}
 R_G\simeq \left(\matrix{1+ \eta_e & 0 & 0 \cr 0 & 1+\eta_\m & 0 \cr 0 & 0
& 1\cr}
  \right), \quad
 \eta_i = 1- \sqrt{\frac{I_i}{I_\tau}} \  (i=e,\ \mu) , \quad
I_i\equiv\exp {\left (\frac{1}{8\pi^2}\int\limits_{\ln{(Mz)}}^{\ln{(M_R)}}
y_i^2\ dt
      \right )}.
  \label{RGE}
  \end{equation}
  \noindent
We transform back  this neutrino mass matrix $R_G \bar M_\n R_G$ into the
basis where the charged lepton mass matrix is the democratic one at the EW
scale as follows:
 \begin{equation}
   F R_G \bar M_\n R_G F^T \simeq {c_\n}\left( \matrix{1+\bar r & \bar r &
\bar r \cr
                            \bar r & 1+\e+\bar r & \bar r \cr
                            \bar r & \bar r & 1+\d+\bar r \cr } \right)
	+ 2 \eta_R {c_\n} \left( \matrix{1 & 0 & 0 \cr
                            0 & 1 & 0 \cr  0 & 0 & 1 \cr } \right) , 
     \quad \bar r = r - \frac{2}{3}\eta_R \ .
 \label{MassR}
 \end{equation}
Here we take $\eta_R\equiv \eta_e\simeq \eta_\mu$, which is a good approximation
 \cite{Haba}.  Its numerical value depends on  $\tan\b$ as:
 $10^{-2}$, $10^{-3}$ and $10^{-4}$ for $\tan \beta=60, \ 10,$ and $1$,
respectively.
 As seen in eq.(\ref{Mass}) and eq.(\ref{MassR}),
   radiative corrections are absorbed into the original parameters
  $r$, $\e$ and $\d$ in the leading order.
 Thus the structure of the mass matrix is stable against radiative 
  corrections although our model leads to nearly degenerate neutrinos.
  Of course, it does not means that the radiative corrections are small.

  We have obtained the neutrino mass matrix with the large mixing angle MSW
solution,
$\sin^2 2\th_\odot=0.65\sim 0.97$ and $\Delta m_{\odot}^2= 10^{-5}\sim
10^{-4}\eV^2$,
   in the $S_{3L}\times S_{3R}$ flavor symmetry.
 The structure of our neutrino mass matrix is found to be 
 stable against radiative corrections. 
 We wait for results in KamLAND experiment as well as new solar neutrino data.
 

\end{document}